\documentclass{elsart} 
\usepackage{epsfig,psfig}
\begin{document}

\emph{Physica A (2005), to appear}.

\begin{frontmatter}
\title{The Reversible Phase Transition
of DNA-Linked Colloidal Gold Assemblies}

\author{Young Sun, Nolan C.\ Harris, and Ching-Hwa Kiang$^*$}
\address{Department of Physics and Astronomy\\
Rice University, Houston, TX\ \ 77005--1892}

\begin{abstract}
We present direct evidence for a reversible phase transition of
DNA-linked colloidal gold assemblies. Transmission electron microscopy
and optical absorption spectroscopy are used to monitor the colloidal
gold phase transition, whose behavior is dominated by DNA
interactions. We use single-stranded DNA-capped colloidal gold that is
linked by complementary linker DNA to form the assemblies. We found that,
compared to free DNA, a sharp melting transition is observed for the
DNA-linked colloidal gold assemblies.  The structure of the assemblies
is non-crystalline, much like a gel phase, consistent with theoretical
predictions. Optical spectra and melting curves provide additional
evidence of gelation of the colloidal system.  The phase transition
and separation are examples of percolation in a dilute solvent.
\end{abstract}

\begin{keyword}
DNA phase transition \sep gold nanoparticle  \sep DNA melting

\PACS 82.39.Pj \sep 87.15.By \sep 87.15.He \sep 87.68.+z \sep 87.15.-v
\end{keyword}

\end{frontmatter}


\section{Introduction}

Colloidal nanoparticles functionalized and linked with single-stranded
DNA exemplify a new class of complex fluids.  Both equilibrium and
nonequilibrium phase transitions of complex particle systems are
of great interest \cite{Frenkel02a,Zia89a}. The interaction between
colloidal particles is controlled by the DNA intermolecular
potentials, which are dominated by hydrogen bonding.  Owing to the
complex yet specific nature of DNA base pairing, the interaction
between colloids in such a system can be precisely controlled and
``tailor made'' to have a specific potential.  Both the number of
components and the strength of the interaction forces can be designed.
A variety of different states of this system have been investigated
theoretically \cite{Frenkel04a,Tkachenko02a}.

DNA melting and hybridization, an important process in DNA replication
and translation, have been studied for decades
\cite{Nelson00a,Zocchi03a,Rudnick03a,Wartell85a,Hwa97a,Fisher66a,Grassberger00a,Breslauer99a}.  
Many of the thermodynamic
properties of free DNA are known, yet DNA interactions in
constrained spaces such as on surfaces, {\em e.g.}\ in microarrays,
are still poorly understood \cite{Magnasco02a}.  The sequence-specific
hybridization properties of DNA have been used for self-assembly of
nanostructures \cite{Mirkin96a,Alivisatos96a}.  The macroscopic
properties of these novel systems can be easily detected and are a
result of the microscopic properties of DNA. Thus, studies of the
phase behavior of these self-assemblies provide valuable information
on fundamental DNA properties \cite{Frenkel04a,Stroud03a,Lekkerkerker02a}.

DNA-linked gold colloids were thought to self-assemble into
crystalline structures \cite{Mirkin96a}; however, it was proposed by
Kiang \cite{Kiang03a} that the structure of the assemblies is
amorphous, much like a percolating cluster.  Calculated optical
spectra are consistent with an amorphous structure
\cite{Stroud03a}.  The phase behavior of this system has also been
investigated theoretically, and it is believed that the system
undergoes a liquid-liquid phase separation, with the dense liquid
phase behaving as a solid (amorphous) gel \cite{Frenkel04a}.  More
experimental studies of the structural phase transition are needed to
unravel the true nature of the phase transition, which is crucial in
understanding the behavior of such multi-component complex fluids.

Direct imaging is a powerful tool to study the structures arising
during the colloidal phase transition \cite{Weitz00a}, 
and the results can be interpreted
unambiguously.  In this paper, we present evidence of a liquid-gel
phase transition of DNA-linked gold colloid assemblies via direct
imaging.  The aggregation and phase transition of the DNA-linked gold
colloids were studied using transmission electron microscopy (TEM) \cite{Kiang01a}
and optical absorption spectroscopy.  The results obtained suggest that
gelation and phase separation occurs at room temperature in the
DNA-linked gold colloid system.

\section{Results and Discussion}

\begin{figure}[htbp]
\begin{center}
\psfig{file=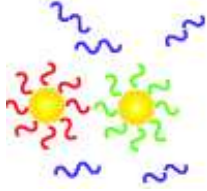,height=1.0in,clip=}
\end{center}
\caption{The basic building block of DNA-linked gold particles.}
\label{fig:basic}
\end{figure}

To prepare colloidal systems with different interactions, we start
with a range of sizes of gold nanoparticles (10 to
40~nm, $<$10\% polydispersity) capped with either 3' or 5'
modified, single-stranded, 12-base DNA.  The basic building block
is illustrated in Fig.~\ref{fig:basic}.  Details of the sample
preparation are described in Ref.~\cite{Kiang03a}.  We added
linker DNA, which caused nanoparticles to form aggregates, and we
studied the melting transition with microscopy and spectroscopy.

\begin{figure}[b]
\begin{center}
\psfig{file=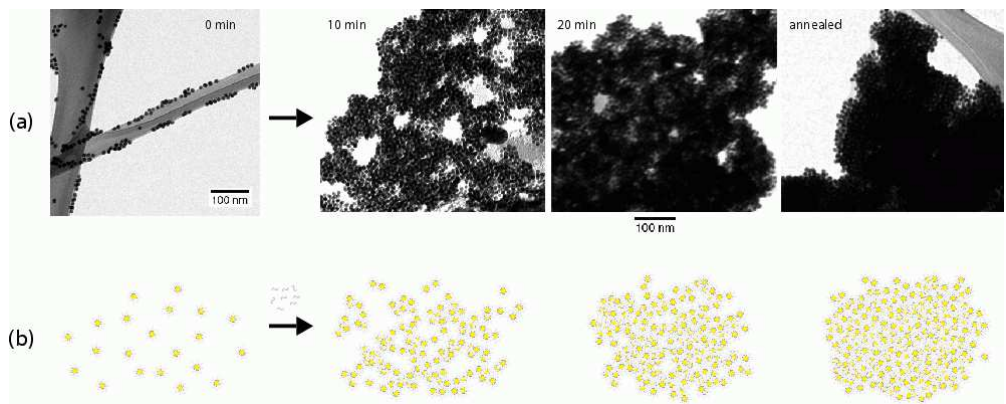,width=0.95\columnwidth,clip=}
\end{center}
\caption{(a) Sequence of transmission electron microscopy images
of 10 nm gold colloids capped with thiol-modified DNA.
The colloids are initially dispersed in solution,
upon adding linker DNA, gold colloids form gel-like porous and amorphous
aggregates.  Phase separation occurs, and eventually the
dense aggregates precipitate out of the solution.
(b) The growth mechanism of DNA-linked gold colloids.
The colloids are initially dispersed in solution.
Upon adding linker DNA, gold colloids form gel-like porous and amorphous
aggregates.  Phase separation occurs, and eventually the
dense aggregates precipitate out of the solution.}
\label{fig:TEM}
\end{figure}

We propose the growth mechanism of DNA-linked gold colloids
(illustrated in Fig.~\ref{fig:TEM}b) to be as follows.
Particles initially are dissolved in the solution. With
the addition of complementary linker DNA, hybridization occurs, and the
particles form a gel-like structure. The connectivity of the
porous structure continues to increase past the percolation threshold,
and eventually the clusters become a dense amorphous structure.
Phase separation occurs and the gel-like aggregates eventually
precipitate out of the solution.

Direct imaging using transmission electron microscopy (TEM)
supports this model.  Fig.~\ref{fig:TEM}a displays a series of
TEM images taken at different aggregation stages.
Before adding linker DNA, the DNA-capped gold particles are
dispersed in the solution.
Upon adding linker DNA, the colloidal gold forms aggregates.
The process is reversed by raising the temperature to above
the transition temperature.

UV-visible absorption spectroscopy is a powerful tool to study the
aggregation and phase transition of the DNA-linked gold colloids
because DNA bases and gold colloids have strong absorption in the UV
region ($\sim$ 260 nm) and the visible light region ($\sim$ 520 nm),
respectively. The extinction coefficient as well as the peak position
are sensitive to the size of the aggregates of gold colloids.
The DNA double helix has a smaller
extinction coefficient than does single-stranded DNA due to
hypocromism \cite{CantorII}.
Thus, both the kinetics of colloidal aggregation and
the temperature-dependent melting transition can be
investigated using UV-visible spectroscopy. All spectra were taken on a
PerkinElmer Lambda 45 spectrophotometer.

\begin{figure}[bt]
\begin{center}
\epsfig{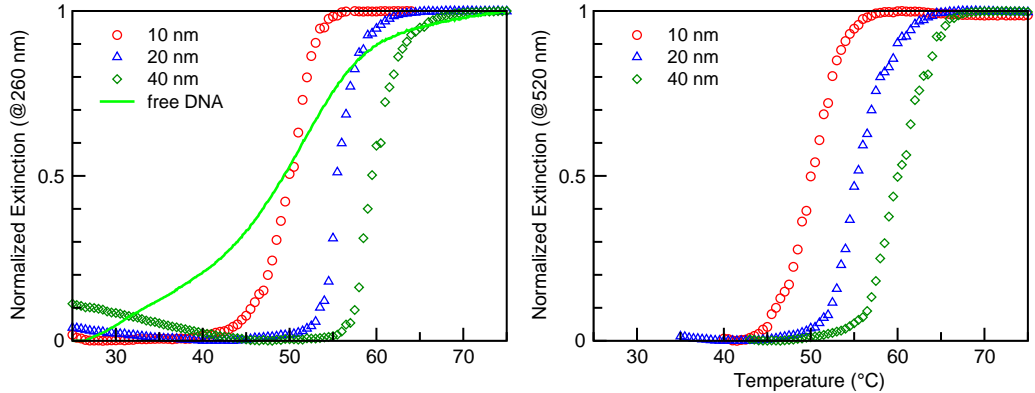}
\end{center}
\caption{Melting curves for $AB$ nanoparticle systems monitored
at (a) 260 nm and (b) 520 nm.
The free DNA melting curve monitored at 260 nm is shown in (a).}
\label{fig:melting}
\end{figure}

Upon adding linker DNA, gold nanoparticles begin to aggregate,
as indicated by the change of the UV absorption.
The aggregation rate at room temperature is faster for
systems with higher $T_m$.
Solutions of DNA-linked gold colloids were allowed to stand at room
temperature for several days for the system to fully aggregate
before melting studies.  We monitor the
absorption intensity at 260 and 520 nm while slowly heating the
DNA-linked gold colloids. The sample was heated by a peltier
temperature controller from 25 to 75 $^\circ$C at a rate of 0.5 $^\circ$C/min.
The 260 and 520 nm melting curves are very similar,
indicating that DNA and gold colloid melting are closely
related. Fig.~\ref{fig:melting} shows the melting curves of 10, 20, and 40 nm gold
colloids with linker DNA.  For comparison, the melting transition
of a free DNA is also shown.  Apparently, the melting transition of
gold-attached DNA is much narrower than that of free DNA.

The melting temperature $T_m$, defined as the temperature at the
midpoint of the absorbency transition \cite{Crothers00a}, is found to
be a function of particle diameter, $D$, and the data are fitted to
\mbox{$T_m(^\circ C)=68-57/\sqrt D$}.  The surface coverage of
thiol-capped DNA bound to gold nanoparticles has been determined via
the fluorescence method to be approximately 160 DNA (12-mer) bound for
a 16~nm diameter particle \cite{Mirkin00c}.  Assuming the number of
DNA bound to gold particles scales with the particle surface area,
$D^2$, there are 50, 190, and 750 DNA on 10~nm, 20~nm, and 40~nm
particles, respectively.  An increasing number of connections between
particles as the gold particle size increases is expected.  Since the
hydrogen bonding energy per DNA pair remain the same, an increased
number of connections effectively increases the enthalpy, $\Delta H$,
between particles and, therefore, raises the melting temperature, $T_m$
\cite{Stroud03a}.

For short DNA (12--14 base pairs), melting and hybridization
can be described by a two-state model as an equilibrium
between single- and double-stranded DNA \cite{Crothers00a},
$$
S + S \rightleftharpoons D.
$$
The melting curve is a slowly varying function of temperature
and can be described by the van't Hoff relationship \cite{CantorII}.
While short, free DNA does not exhibit a phase transition,
DNA bound to nanoparticles aggregate to form networks that have
a definite phase transition, and the melting curves cannot be described
with the van't Hoff relationship.

\begin{figure}[!b]
\begin{center}
\psfig{file=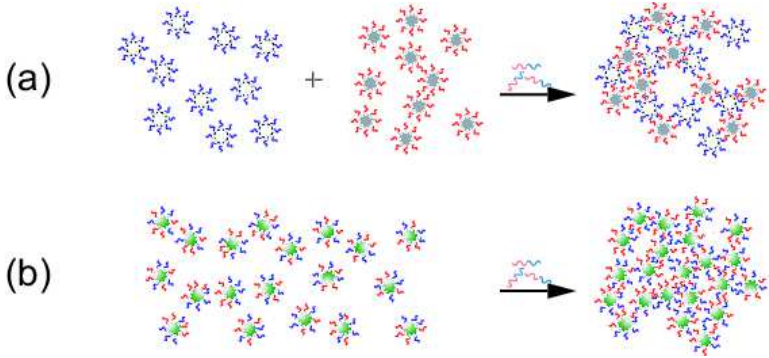,width=4in,clip=}
\end{center}
\caption{Compositions of the $AA$ versus $AB$ systems.
(a) In the $AB$ system, each particle is covered with 
with either $DNA1$ or $DNA2$ to create probes $A$ and $B$, respectively.
$DNA1$ and $DNA2$ are not complementary, and the linker DNA is 
complementary to $DNA1$ and $DNA2$.  As a
result, probe $A$ binds exclusively with probe $B$ upon the
introduction of a target DNA sequence. 
(b) In the $AA$ system, each
particle is covered with both $DNA1$ and $DNA2$ at a 1:1 ratio, 
resulting in only one type of probe ($A$). 
Hence each particle can bind with any other particle in the solution.}
\label{fig:aavab}
\end{figure}

The DNA-linked gold nanoparticle assemblies have several unusual
features, including a sharp melting transition compared to
that of corresponding free DNA, and the melting temperature, $T_m$,
dependence on colloid size \cite{Kiang03a,Kiang05a,Kiang05b}.  
Recently, simulations based on the bond
percolation model \cite{Stroud03a,stroud03b} agree
qualitatively with the experimental optical spectra of DNA-capped
gold colloids at the melting transition.
It has been suggested that at percolation,
$
[1-p(T_c)]^{N_s/z} = 1 - p_c,
$
where $N_s, z, p_c, T_c$ are number of single-stranded
DNA on each gold particle, number of nearest neighbors per gold particle,
bond percolation threshold, and melting temperature,
respectively \cite{Stroud03a}.
The fraction of single-stranded DNA that form links
is $p(T)$ and is temperature dependent.
The melting transition occurs when the fraction
of links falls below the percolation threshold.  The calculated optical
properties are found to change dramatically when this threshold is
passed.  Moreover, the size dependence of melting temperature can be
explained in terms of the effect on link fraction.
The number of DNA on each particle, $N_s$, increases
with increasing particle size, $D$, with the relation $N_s \propto D^2$.
The simulation confirmed that the melting temperature
dependence with particle size.
Therefore, the percolation model can qualitatively explain the
experimental findings.

The experimental results used for comparison, however,
have two kinds of particles, denoted as an $AB$ system since
each particle is either covered with $DNA1$ (denoted as particle $A$)
or $DNA2$ (denoted as particle $B$), but not both, and particle 
$A$ can only be linked to particle $B$.  Fig.~\ref{fig:aavab}a illustrates
such a binary system.  
In most simulations, the system is composed of only one
kind of particle, and each particle can be linked with any other
particle.  To this end, 
we studied an $AA$ system, where each particle is covered with
both $DNA1$ and $DNA2$.  This results in only one type of nanoparticle,
and each particle can be linked with any other particle in the system, 
as shown in Fig.~\ref{fig:aavab}b.

\begin{figure}[bt]
\begin{center}
\epsfig{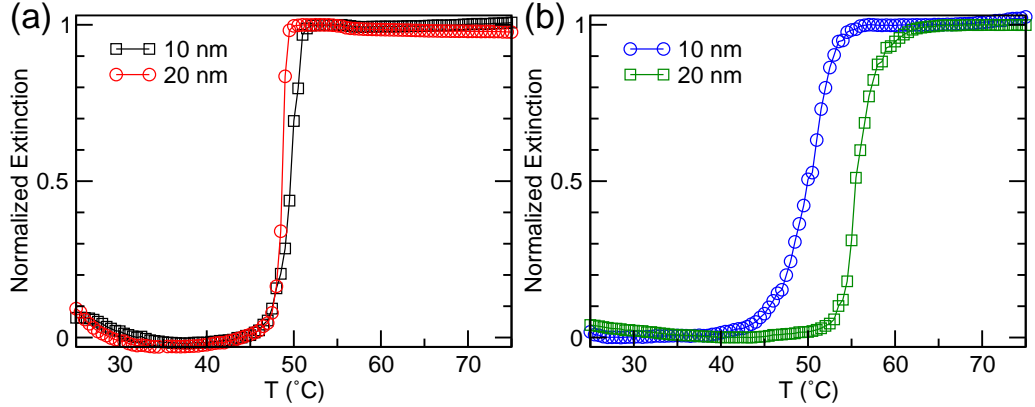}
\end{center}
\caption{Melting curves for (a) $AA$ and (b) $AB$ nanoparticle systems
monitored at 260~nm. 
In the $AB$ (binary particle) system, particle $A$ can only bind
with particle $B$ and {\em vice versa}, whereas in the $AA$ system
there is only one type of particle.
The melting transition width is different and the melting
temperature scales differently for the $AA$ and $AB$ systems.}
\label{fig:aaabmelting}
\end{figure}

Melting curves for $AA$ and $AB$ systems are shown in 
Fig.~\ref{fig:aaabmelting}.  We noticed that the width of the melting transition 
in the $AA$ system is much narrower than that of the $AB$ system.
The scaling of melting temperature with particle size is also different
for $AA$ and $AB$ systems.  
Unlike the $AB$ system,
there are $DNA1$ and $DNA2$ on the same particle in the $AA$ system.
Since $DNA1$ only connects with $DNA2$, the connections in the $AA$ system
is different from that in the $AB$.
More experimental data on the melting curves of $AA$ systems of different
particle sizes are necessary to elucidate how the trend in melting temperature
changes from the $AB$ to the $AA$ systems.

As Lukatsky and Frenkel \cite{Frenkel04a} recently proposed, the
equilibrium phase behavior of DNA-linked colloidal assemblies is
dominated by the temperature-dependent binding free energy of a
double-stranded DNA connecting a pair of gold colloidal
particles.  Depending on the strength of the interaction, the system
can be in a homogeneous state, or separate into two
coexisting phases. Their calculation shows that there is a
liquid-liquid phase separation in DNA-linked gold nanoparticle
assemblies. The origin of the sharp phase transition is the entropic
cooperativity of DNA-nanoparticle network. Upon cooling, the system
undergoes liquid-liquid phase separation.  The dense liquid phase is
strongly cross-linked and behaves as a solid gel.

\begin{figure}[bt]
\begin{center}
\psfig{file=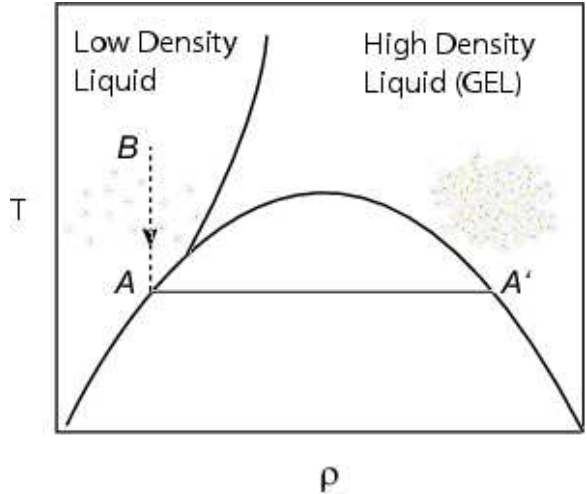,width=3in,clip=}
\end{center}
\caption{Proposed phase diagram of a DNA-linked colloidal gold system.
The gelation in the presence of solvent results in phase separation
below the transition temperature.  The experiment follows the dashed
line to the liquid/gel phase separation.}
\label{fig:phase}
\end{figure}

In general, the percolation model is a crude
representation of any gelation processes \cite{deGennes79a}.
In practice, the DNA-capped gold colloids are mixed
with a solvent, and this can be considered as
``dilution effect,'' {\em i.e.} gelation in dilute solution.
A gelation process in the presence of solvent always brings a
trend toward phase separation of the gelating species,
and the phase diagram is illustrated in Fig.~\ref{fig:phase}.
However, the critical exponents observed in our system may
be of the percolation type, as predicted by de~Gennes \cite{deGennes79a},
and supported by simulations \cite{Stroud03a}.
For our systems, we did observe phase separation, as predicted by
de~Gennes \cite{deGennes79a}.  TEM images give direct evidence of the formation of
the gel phase, and support the expectation of phase transition and separation.
As shown in Fig.~\ref{fig:TEM}, upon adding linker DNA, gold colloids form
gel-like porous and amorphous aggregates.  Phase separation occurs and
eventually the dense aggregates precipitate out of the solution,
and the solution eventually became clear.

\section{Summary}

We have presented direct evidence of a liquid-gel
phase transition of the DNA-linked gold colloidal assemblies.
Compared to free DNA, a sharp melting transition is observed
for this system. The formation of DNA-linked gold colloids
and the sharpness of the melting transition resemble the
phase behavior of gelation in dilute solution phenomena.
We have shown that the scaling property of the binary system $AB$, where two types
of particles exists, is different from that of $AA$, where
all the particles are of the same type.  
The results shown here indicate that the DNA-linked
gold nanoparticles represent a new class of complex fluids. 

$^*$To whom correspondence should be addressed,
email: chkiang@rice.edu.

\bibliography{aa}

\clearpage

\clearpage

\end{document}